\documentclass[preprint,iop]{emulateapj}

\usepackage{mathtools}
\usepackage{epsfig}
\usepackage{amssymb,amsmath}
\usepackage{placeins}
\usepackage{natbib}
\usepackage{wasysym} 
\usepackage{verbatim}
\newcommand{\argmax}{\operatornamewithlimits{arg\ max}}

\newcommand{\Kepler}{\textit{Kepler}\ }
\newcommand{\impurity}{\mathbf{I}}
\newcommand{\corrthreshold}{\theta_{\rho}}

\slugcomment{Published 2015 ApJ 806 6 doi:10.1088/0004-637X/806/1/6}

\shorttitle{Automatic Classification of Transit Candidates}
\shortauthors{McCauliff et al.}

\begin{document}

\title{Automatic Classification of Kepler Planetary Transit Candidates}
\author{Sean D. McCauliff\altaffilmark{3},
Jon~M.~Jenkins\altaffilmark{1},
Joseph Catanzarite\altaffilmark{2}, 
Christopher J. Burke\altaffilmark{2},
Jeffrey L. Coughlin\altaffilmark{2},
Joseph D. Twicken\altaffilmark{2}, 
Peter Tenenbaum\altaffilmark{2}, 
Shawn Seader\altaffilmark{2}, 
Jie Li\altaffilmark{2}, 
Miles Cote\altaffilmark{1}
}

\email{sean.d.mccauliff@nasa.gov}
\altaffiltext{1}{NASA Ames Research Center, Moffett Field, CA 94035, USA}
\altaffiltext{2}{SETI Institute/NASA Ames Research Center, Moffett Field, CA
94035, USA}
\altaffiltext{3}{Wyle/NASA Ames Research Center,
Moffett Field, CA 94035, USA}


\begin{abstract}
In the first three years of operation the \Kepler mission found  3,697 planet candidates from a set of 18,406 transit-like features detected on over 200,000 distinct stars. Vetting candidate signals manually by inspecting light curves and other diagnostic information is a labor intensive effort.  Additionally, this  classification methodology does not yield any information about the quality of planet candidates; all candidates are as credible as any other candidate.  The torrent of exoplanet discoveries will continue after \Kepler as there will be a number of exoplanet surveys that have an even broader search area. This paper presents the application of machine-learning techniques to the classification of exoplanet transit-like signals present in the \Kepler light curve data.  Transit-like detections are transformed into a uniform set of real-numbered attributes, the most important of which are described in this paper.  Each of the known transit-like detections is assigned a class of planet candidate; astrophysical false positive; or systematic, instrumental noise.  We use a random forest algorithm to learn the mapping from attributes to classes on this training set.  The random forest algorithm has been used previously to classify variable stars; this is the first time it has been used for exoplanet classification.  We are able to achieve an overall error rate of 5.85\% and an error rate for classifying exoplanets candidates of 2.81\%.  

\end{abstract}

\keywords{astronomical databases: miscellaneous, binaries: eclipsing, catalogs, methods: statistical, planets and satellites: detection, techniques: photometric}

\section{Introduction}
\Kepler is a single instrument spacecraft that maintains its pointing at the designated field almost constantly in order to collect the most contiguous and long-running, photometric, time series possible\footnote{While operating, \Kepler can simultaneously observe approximately 170,000 stars.  Data gaps on the order of days result from a monthly data down link and a quarterly 90-degree roll used to maintain the orientation of the solar panels.  This roll causes the vast majority of stars to be observed by a different CCD every three months with a one year cycle.}.  The primary mission of \Kepler was to detect large numbers of transiting exoplanets.  The ultimate goal of the primary mission was to characterize the frequency of exoplanets with respect to diameter, orbital period and host star.  Manual classification of the findings of \Kepler has proven very time consuming and provides only coarse ranking of unclassified data.  New, space-based, transit photometry missions such as K2\citep{howell2014K2}, TESS\citep{ricker2014Tess}, and PLATO 2.0\citep{rauer2014Plato} will also be plagued with the same embarrassment of riches that \Kepler has faced and will require some level of automation.

Using a machine learning approach we can speed this process and provide a more continuous ranking of planet candidates. The technique presented in this paper has already been used to assist in finding new, habitable-zone planets \citep{torres2015Validation}.

A threshold crossing event (TCE) is a sequence of significant, periodic, planet transit-like features in the light curve of a target star.  TCEs are subjected to a vetting process performed by the \Kepler TCE Review Team (TCERT).   An initial stage of vetting, known as triage, partitions the set of TCEs into problematic light curves that have instrumental noise and \Kepler objects of interest.  A \Kepler object of interest is a TCE that contains convincing transit-like features that do not present any obvious evidence that the TCE was generated from non-transiting phenomena (e.g. instrumental noise). These potential objects are further scrutinized with \Kepler data in later stages of vetting.   These later stages seek to find evidence that the signal results from an eclipsing binary star or more subtle forms of instrumental noise.  TCEs that do not present such evidence are disposed as planet candidates (PC).   This process is performed by individuals manually inspecting light curves and detection statistics.  The vetting process is described in greater detail in these KOI catalog papers \citet{batalha2013Planetary,burke2014Koi,rowe2015Koi,mullally2015planetary}.

We automate this process of classifying TCEs into one of three classes: planet candidate (PC), astrophysical false positives (AFP) and non-transiting phenomena (NTP). PCs are confirmed as planets, statistically validated as planets, or determined to be a planet candidate by the TCERT.   AFPs are those TCEs that have been shown to be eclipsing binary stars or have shown evidence that the transiting object being detected is not located around the target star. The class NTP are those TCEs that failed triage.

We use a machine learning algorithm known as the random forest \citep{breiman2001Random} to find a function that maps attributes produced by the \Kepler Pipeline for each TCE to a classification of PC, AFP or NTP. This classification function is purely based on the statistical distributions of the attributes for each TCE (e.g. SNR of the transit model fit) with respect to TCEs classified in the manual vetting process.  These algorithms do not attempt to physically model the process of planet transits beyond what is already present in the TCE attributes. Random forests have been applied to a variety of classification problems, perhaps closest to this problem is that of classifying variable stars \cite{dubath2011Random} and \cite{richards2011machine}.

This paper is organized as follows.  In section \ref{sec:pipeline} we give an overview of the relevant portions of the \Kepler pipeline that generate the inputs to the random forest.  Section \ref{sec:randomForests} introduces notation and describes the rationale behind the random forest algorithm. In section \ref{sec:dataSet} we discuss the provenance of our data and the source of our known class labels.  Section \ref{sec:attributes} talks about how we have pruned our set of attributes and which attributes turned out to be the most important.  Section \ref{sec:results} contains an analysis of the  classification performance and a comparison to reference classification algorithms on the training data set.  Section \ref{sec:application} discusses issues involving how the automated classification would be applied to an exoplanet transit survey.  A survey for transiting planets will initially discover larger, shorter period planets before discovering smaller, longer period planets so that the training set will lag in period behind the unknown data. Section \ref{sec:longer} shows what happens to classfication performance when this happens.  Sections \ref{sec:later} and \ref{sec:betterPriors} deal with testing and adjusting priors.  Finally, section \ref{sec:conclusion} talks about future work we may perform to assist in discovering additional transiting exoplanets as well as to improve the performance of the classifier.

\section{\Kepler Pipeline}
\label{sec:pipeline}
\subsection{\Kepler light curves}
A single sample of photometric data is composed of a set of co-added 6 second integrations with a readout time of 0.5 seconds whose durations total 29.4 minutes\footnote{29.4 minute co-adds are known as long cadences.  Some stars observed at a higher sampling rate of approximately one minute co-adds}.  Co-added pixel values are downloaded from the spacecraft with the remainder of the processing performed on Earth.  After calibration, light curves are produced using simple aperture photometry and subjected to cotrending to remove systematic noise \citep{smith2012regularMAP,stumpe2012pdcArch}.

\subsection{Transiting Planet Search}
\label{sec:TPS}
The \Kepler Pipeline \citep{jenkins2010overview} is a data reduction pipeline used for translating the \Kepler raw pixel data into potential transiting planet detections.   In particular, this paper concerns itself with the last two modules of the pipeline; those that identify TCEs and their subsequent transit model fitting.  The Transiting Planet Search \citep{jenkins2010transiting} module takes as input the systematic error-corrected light curve for a star; searches a parameter space of possible transit signatures; and outputs a TCE or says that one does not exist on the target star.  This produces a smaller list of stars that have TCEs.  This subset of target stars is then given to the ''data validation'' module.  Data validation fits this initial TCE to a transit model using the geometric transit model described by \citet{mandel2002analytic} with the limb-darkening coefficients of \citet{claret2011Limb}.  Data Validation then gaps the transit signature from the light curve and uses the Transiting Planet Search to find additional TCEs on the same target star. This process repeats until no more TCEs are found on a star or a processing timeout has been reached.

The Transiting Planet Search (TPS) algorithm detects transit-like features in light curves by applying a noise-compensating, wavelet-based matched filter\citep{jenkins2002Impact}.  TPS characterizes the power spectral density (PSD) of the observation noise as a function of time to implement a whitening filter in the wavelet domain. The trial transit pulse is whitened and correlated against the whitened flux time series. Features with correlations above the threshold of $7.1 \sigma$ are flagged as potential threshold crossing events and subjected to additional tests in TPS to guard against false alarms \citep{seader2013chi2}.

  The algorithm searches a parameter space of varying transit duration ($ D  \in $ \{1.5, 2.0, 2.5, 3.0, 3.5, 4.5, 5.0, 6.0, 7.5, 9.0, 10.5, 12.0, 12.5, 15.0\} hours).  This produces a Single Event Static (SES) time series that is the significance of the detection of the reference transit pulse centered on that particular time for each D.  This is computed as $ SES(t) = \mathbb{N}(t)/\sqrt{\mathbb{D}(t)} $, where $ \mathbb{N}(t) $ is the correlation time series; that is, how well the reference transit pulse correlates with the light curve.  The quantity, $\sqrt{\mathbb{D}(t)}  $, is the expected signal to noise ratio of a signal that exactly matches the template pulse.  The Combined Differential Photometric Precision is $ 1E6 / \sqrt{\mathbb{D}(t)}$ ppm and encapsulates the effect of the observation noise on the detectability of the reference transit pulse as a function of time.
  
A Multiple Event Statistic (MES) is constructed that characterizes a significant detection in a search over varying orbital periods $p$ and epochs (phase) $ t_0 $ by folding $\mathbb{N}(n) $ and $ \mathbb{D}(n)$. A $ MES > 7.1 \sigma $ may produce a TCE if it also passes additional statistical tests described by \citet{seader2013chi2}.  If several permutations of $ (t_0, p, D)$  would produce a TCE, then TPS only reports the event with the maximum $MES $ as the TCE for the target star.  Diagnostics such as MES and SES are the basis of some of the attributes used in the training set (see section \ref{sec:mostImportantAttributes}).  The minimum MES, which can be negative if the light curve is anti-correlated with the reference transit pulse, is also computed. This is referred to as the $MES_{min}$.  An event with a negative $MES_{min}$ would look like a repeated brightening above the median flux.

In addition to detecting PCs and AFPs, Transiting Planet Search often generates a TCE for non-transiting phenomina.  NTPs are caused by instrumental, systematic noise \citep{caldwell2010Instrument}, and stars with high variability such as ``heartbeat'' stars \citep{thompson2012Class}.  Many NTPs have periods with approximately one year in duration as the electronics associated with the instrumental signatures rotates back into the same observing field.  Unfortunately, this part of parameter space corresponds to the habitable zone of Sun-like stars.

\subsection{Data Validation}
Data validation  \citep{wu2010Data} is a set of algorithms that perform tests to determine the suitability of the TCE as a planet candidate.  In particular, data validation fits the transiting portion of the light curve to a transit model using the geometric transit model described by \citet{mandel2002analytic} with the limb-darkening coefficients of \citet{claret2011Limb} and constructs a set of diagnostic tests to help determine the nature of the TCE.  Data validation also attempts to locate the actual source of the transits \citep{bryson2013identification}, in order to determine if the actual transit signal is being induced by a source offset from the target star.  This can be caused by background eclipsing binaries, optical ghosts of bright stars with periodic signatures, or other unresolved contamination \citep{coughlin2014Contamination}.  Data validation constructs difference images between the in-transit and out-of-transit images for the target.  The centroid of the difference image is then the center of the transit source.  If the difference image is significantly offset from the out-of-transit centroid, then a background eclipsing binary may have produced the transit-like signal.  Using the transit centroid offset from the position of the target star in the \Kepler Input Catalog (KIC) or the out-of-transit centroid offset turns out to be useful for detecting AFPs (section \ref{sec:mostImportantAttributes}).  
 
\FloatBarrier

\section {Random Forests}
\label{sec:randomForests}
\subsection{Introduction}
\begin{figure}
\epsfig{file=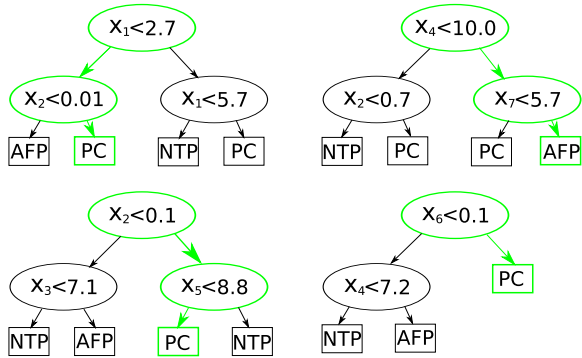,width=3.0in}
\caption{A hypothetical, simplified random forest. Left branches are taken if the inequality is satisfied otherwise the right branch is taken.  A TCE with the attribute vector of $x = $ ( 2, 0.2, 4, 11, 5, 0.3, 6.1) is classified.  Bold, green lines indicate paths taken in the decision trees.  The individual decision trees have voted \{ PC, AFP, PC, PC \} yielding a classification of PC for the forest as a whole. 
 \label{fig:exampleRandomForest}}
\end{figure}

A random forest itself is an ensemble of decision trees.  Each tree votes for the classification of an unknown object based on a vector of attributes (real valued random variables) that describe the object in question.  The classification the forest assigns to an object is the plurality vote of the random forest.  A decision tree is a directed graph where each internal vertex in the graph is a binary test; in this case, the tests are inequalities.  If the test is true then one branch is taken else the other branch is taken.  Leaf vertices indicate classifications.  The trees in the random forest are constructed in such a way that attributes under test are randomly selected and so each tree makes a different set of tests.  Figure \ref{fig:exampleRandomForest} shows a simplified, hypothetical random forest.

\subsection{Notation}
 $ \mathbf{X} $ is the set of random variables that describe a TCE.  Sometimes we refer to $\mathbf{X}$ as the attributes of the TCEs.  When we need to refer to a particular random variable without naming it, we use $\mathbf{x}_i$ or some other subscript.  So the set of attributes of a TCE could be referred to as $\mathbf{X} = \{ \mathbf{x}_1, \ldots, \mathbf{x}_A\}$. Where $A$ is the number of attributes or  $ |\mathbf{X} |$.  The instantiation of the set of all the random variables for all the TCEs under consideration is $X$.  When referring to the set of instantiations for a particular TCE we use $x$.  The constant, $s_i$, is used to split $X$ into two different subsets.  The subscript $i$ refers to a split made on the ith attribute.  $\mathbf{Y}$ refers to the set of all classes to which a TCE might belong; these are sometimes known as labels.  A particular label is denoted $\mathbf{y}_i \in \mathbf{Y}$ and there are $C = |\mathbf{Y}|$ distinct classes.  For this classification problem, $\mathbf{Y} = \{\mathbf{y}_1, \mathbf{y}_2, \mathbf{y}_3\} = \{ PC, AFP, NTP\} $.  $Y$ refers to the instantiation of all the labels for all the TCEs under consideration.  $y$ is the instantiation for a particular TCE (usually in the training set). The set $\mathcal{L}$ is a set of instantiations of all the attributes $ X $ and their labels $ Y $;  that is, attributes that describe all the TCEs and the TCEs' labels. $\mathcal{L}$ is also known as the training set.  The size of the training set is $ N = | \mathcal{L}| $; the number of labeled TCEs.
 
When a split is generated one sub-training set contains all of the TCEs with an attribute less than the split, $ \mathcal{L}_l = \mathcal{L}_{\mathbf{x}_i < s} $. The other sub-training set is composed of all the TCEs greater than or equal to the split $ \mathcal{L}_r = \mathcal{L}_{\mathbf{x}_i >= s} $.  The number of TCEs in each of these respective sub-training sets are then $N_l $ and $ N_r $.

The purpose of classification is to assign an estimated label, $\hat{y}$, given an instantiation of attributes $x$.  A classification function, $h$, generates a prediction for a set of attribute instantiations, $ h: x \to \hat{y} $. 

\subsection{Decision Trees}
As an example, equation \ref{eqn:simpleRule} is a simple, makeshift, classification function. The training set consists of only two attributes to describe a TCE and two splits.
\begin{multline}
\label{eqn:simpleRule}
    h(MES_{min} / MES_{max}, \Delta_{centroid}) =  \\
\begin{cases}
    \text{classify TCE as NTP}, \text{if}\: \frac{MES_{max}}{MES_{min}} < s_1 \\
    \begin{cases}
      \text{classify as PC},& \text{if}\: \Delta_{centroid} < s_2 \\
      \text{classify as AFP},& \text{otherwise} \\
    \end{cases},
\end{cases}
\end{multline}

$\Delta_{centroid}$ is the difference between the star's \Kepler Input Catalog position and its mean out-of-transit centroid in arc seconds.  $MES_{min}$ is the minimum, multiple event statistic discovered by the TPS, indicating repeated events above the baseline stellar flux.  $MES_{max}$ is the highest multiple event statistic discovered by TPS.  A TCE representing a strong transit detection should have a high ratio of these two attributes.  This is attribute \ref{attr:allPulseMaxMesMinMesRatio} in section \ref{sec:mostImportantAttributes}.  TCEs caused by background eclipsing binaries should show an offset from their catalog position, so a high $\Delta_{centroid}$ value is cause to classify a TCE as an AFP.  

Optimizing the parameters $ s_1, s_2$ for the minimal misclassification rate of $ h $ yields a misclassification error rate of $ 0.0886 $.  The misclassification error rate is simply the number of incorrect classifications divided by the size of the training set ($N$).  A logical next step to reduce the error rate would be to consider additional attributes that would discriminate TCEs.  The Classification and Regression Tree (CART) algorithm \citep{breiman1984CART} can construct a recursive nesting of conditional tests of arbitrary depth.  These are sometimes visualized as a tree of conditionals called a decision tree.  When building a decision tree, CART minimizes the following function for all splits $s_i$ under consideration:

\begin{equation}\label{eqn:impurityDelta}
\Delta \impurity(\mathcal{L}, s)  = 
      N \impurity(\mathcal{L}) 
    - N_l   \impurity(\mathcal{L}_l) 
    - N_r   \impurity(\mathcal{L}_r).
\end{equation} 

$ P(\mathbf{y}_i) $ is the frequency of class $\mathbf{y}_i$ in $\mathcal{L}$.  Function $\impurity$ is an impurity function (  $ \impurity : \mathcal{L} \to  \mathbb{R} $ )   chosen so that it is maximum when all $ P(\mathbf{y}_i) $ are approximately equal and zero when any $ P(\mathbf{y}_i) = 1 $.  $\mathcal{L}_l $ is the subset of $ \mathcal{L} $ consisting of the training examples where attribute $ \mathbf{x}_i $ is less than the threshold $s$.  The members of  $\mathcal{L}_l$ depends only on the ordering of the attributes relative to the split.  This implies attribute transformations that are class independent and do not change the ordering of training examples will result in the same $\mathcal{L}_l$.  For example, $log\, \mathbf{x}_i$ would not change the results of any splits.  When CART is selecting a split for the training set, it iterates over all $\mathbf{x}_i$ to find the attribute with the largest $\Delta \impurity$.

One possible definition for $\impurity$ is an information function:
\begin{equation}\label{eqn:information}
\impurity_Z(\mathcal{L}) = - \sum_{i=1}^{C} P(y_i|\mathcal{L}) log_2[P(y_i|\mathcal{L})],
\end{equation}
where $P(y_i|\mathcal{L})$ is the probability of a given class, $y_i$ given the training set $\mathcal{L}$.  So that when a split is selected using equation \ref{eqn:impurityDelta} we are maximizing the information loss of the split.  A perfect split would result in the members of $\mathcal{L}_l$ and $\mathcal{L}_r$,   being of the same class and therefore all information would have been lost since $\impurity_Z( \mathcal{L}_l) = 0 $ and $ \impurity_Z(\mathcal{L}_r) = 0$.

\begin{figure}
\epsfig{file=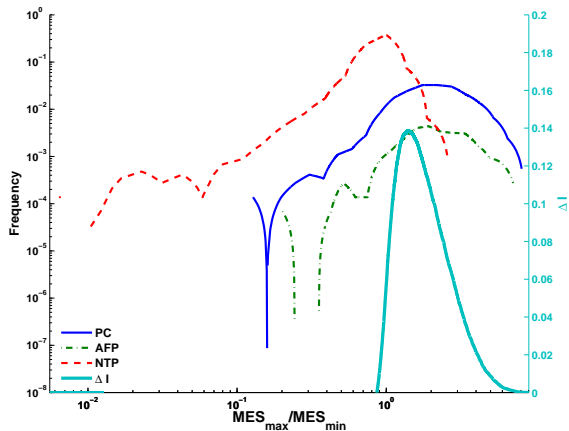,width=3.0in}
\caption{$\Delta \impurity$ computed using impurity function $\impurity_G$for $ MES_{max}/MES_{min} $ over the entire training set.  We would split the training set at the maximum of $\Delta \impurity $.
 \label{fig:giniDelta}}
\end{figure}

In practice the Gini impurity function:
\begin{equation}\label{eqn:giniImpurityFunction}
\impurity_G(\mathcal{L}) = \sum_{i=1}^{C} P(y_i|\mathcal{L}) [ 1 - P(y_i|\mathcal{L})] 
\end{equation}
$\impurity_G$, is used since it is faster to compute and usually results in the same split.  $\impurity_G$ is the sum of the misclassification probabilities.  That is, the probability of guessing class $\mathbf{y}_i$ for some example in $\mathcal{L}$, when the label is actually some other class.  \citet{breiman1984CART} discusses other impurity functions.  Figure \ref{fig:giniDelta} gives an example of $\Delta \impurity$ evaluated at every possible split.  Splitting the training set occurs between the attribute values with the maximum $\Delta \impurity$.

Splitting the training set happens recursively so that $\mathcal{L}_l$ and $\mathcal{L}_r$ are then the training set for the next level of the decision tree.  Tree construction ends when a split would result in a training set with instances of only a single class; impurity has fallen to zero.  This generates the leaf vertex that assigns a class $ \hat{y} $ to $x$ when the tree is evaluated.

\subsection{Random Forest Description}
\label{sec:randomForestDescription}
The forest is a set of base, decision tree classifiers $\{h_1, \ldots, h_k, \ldots, h_{N_{tree}}\}$.  Each tree, $ h_k(\mathbf{x}) $, is trained on $\mathcal{L}_k$ which is a bootstrap replica of $\mathcal{L}$ known as a bag. Each bag is about $2/3$ the size of the training data set.  The observations not contained in the bag, $\overline{\mathcal{L}_k}$, are known as the out-of-bag data . The number of trees $ N_{tree} $ is one of two tunable parameters for this algorithm.  For this problem we use $ N_{tree} = 15,000 $.  
Each tree's misclassification error can be evaluated on its out-of-bag data to provide an unbiased estimate of the misclassification error of the random forest.  Unlike other classification algorithms, the random forest allows us to estimate the misclassification error rate without having to resort to cross-validation to eliminate overfitting.   Section \ref{sec:importance} discusses how this out-of-bag error can be used to estimate the importance of various attributes.

Unlike any individual decision tree the random forest has the nice property of having bounded error.  That is, the overall misclassification error converges to this error bound.  In order to prove this, \citet{breiman2001Random}, makes the assumptions that individual trees make their decisions independently of other trees.  When a split is made during the construction of $h_k$, a random subset of attributes are considered for the split rather than all attributes.  $m_{try}$ is the number of attributes to select randomly during a split.   These two techniques (i.e. bootstrap sampling and choosing a random subset of attributes at each split) break the correlation among the decision trees in the forest so that they approach the ideal of independent classifiers.   

The random forest's error is bounded by the equation,
\begin{equation}
\label{eqn:bounds}
 Err_{RF} \le \overline{\rho}(1-g^2)/g^2.
\end{equation}
Where $g$ is the strength of the the random forest.  This is the average difference between the proportion of votes for the correct class and the maximum proportion of votes for any other class over all of $\mathcal{L}$.  $\overline{\rho}$ is the correlation among all the pairs of trees of the random forest over all of $\mathcal{L}$ after function

\begin{multline}
\label{eqn:R}
    R(k) = \\
\quad    \begin{cases}
      1,  h_k(x)  = y \\
      -1, h_k(x)  = \text{the most common misclassification of x}\\
      0,  otherwise
    \end{cases},
\end{multline}
has been applied to the classification of each tree on the training data.

The error bounds of the random forest then depend on how correlated the decisions of the trees are with each other and how well the forest as a whole separates between the correct class and the next best classification.  \citet{breiman2001Random} contains a proof for convergence of the random forest to these error bounds.

Equations \ref{eqn:bounds} and \ref{eqn:R} imply that correlated attributes also produce correlated errors which can cause a higher generalization error.  In section \ref{sec:removeCorrelation} we remove correlated attributes.

\FloatBarrier

\section{Training Set and Unknown Set}
\label{sec:dataSet}
\subsection{Introduction}
The TCE catalog contains the ephemerides of all the TCEs under consideration.  This is separated into known ephemerides, which constitute the training set, and unknown ephemerides.  Labeling of the TCEs naturally falls out of this process.

\subsection{TCE Catalog}
\label{sec:tceCatalog}
The TCE catalog used is from the first twelve quarters of \Kepler data, described by \citet{tenenbaum2013Detection}.  This TCE catalog contains a total of 18,407 TCEs.  The TCE ephemerides are correlated with the ephemerides of known planets, planet candidates and astrophysical false positives from the NASA Exoplanet Archive\footnote{\url{http://exoplanetarchive.ipac.caltech.edu}}.   We use the cumulative \Kepler object of interest catalog that was available on the date 2013 June 25 for the training data ephemerides.  An older KOI catalog is used purposefully in order to examine more closely the generalization of the random forest section \ref{sec:later}.

In this particular run of the \Kepler pipeline, 2,126 target stars that generate TCEs were excluded from consideration. The majority of these excluded stars are known eclipsing binaries from the \citet{prvsa2011Kepler} catalog.  Planets that have very short periods and relatively deep transits are identified as harmonics and are removed by TPS (e.g. Kepler-10b \citep{batalha2013Kepler}).  These are not accounted for in the TCE catalog since they are not found by the \Kepler pipeline.  Planets with large transit timing variations and planets whose existence can only be inferred by transit timing variations (e.g. Kepler-19b,c \citep{ballard2011Kepler}) are also not in the TCE catalog.  There are some cases where TPS has identified multiple TCEs for a single eclipsing binary.  When this occurs, the TCEs usually have a period that is an integer multiple of an accompanying TCE on the same  target star.  No attempt is made to match these TCEs to the source eclipsing binary and so they are considered unlabeled and not used in the training set.

\subsection{Non-Transiting Phenomena}
For quarter one through quarter twelve of the \Kepler data, the TCERT reviewed all of the TCEs that were not caused by previously known \Kepler objects of interest.  Each TCE was examined by two or more people (vetters) to see if it had one of the following problems: stellar variability, instrumental effects, or anomalous behavior.  A TCE is considered a product of stellar variability if it shows sinusoidal activity especially outside of the primary transit event.  Stellar variability with periods longer than the transit durations is not of concern since this is removed from the light curve before searching for transits.  Instrumental noise is defined as a TCE exhibiting a transit-like signature apparently caused by spacecraft systematics, such as those induced by periodic spacecraft operations and uncorrected transient events such as radiation damage and cosmic ray hits.  In some cases anomalous TCEs occur if there was insufficient data to make a classification, or if the transit did not appear at the expected position in the light curve.  This step in the process of TCE vetting is known as triage.

A TCE in the training set is identified as an NTP if the vetters where unanimous in their decision to reject the TCE as a new \Kepler object of interest.  That is every vetter needed to identify one of the problems in the paragraph above. This yields 11,304 TCEs that are in the class NTP.  Since these were TCEs identified from the pipeline run, their ephemerides did not require matching against a known catalog.

More information about the TCERT process can be found in \citet{batalha2013Planetary,burke2014Koi,rowe2015Koi}.

\subsection{Planet Candidates and Astrophysical False Positives}
Our training set for class PC and AFP are derived from the public \Kepler objects of interest catalog maintained by the NASA Exoplanet Archive.  This is the union of many \Kepler Object of Interest catalogs, described in more detail in \citet{borucki2011Characteristics,batalha2013Planetary,burke2014Koi,rowe2015Koi}.  A \Kepler object of interest can be a confirmed or validated exoplanet (confirmed or validated by one of several methods), a planet candidate (PC), an astrophysical false positive (AFP), or undispositioned. A PC has a convincing transit-like signature in the light curve for the target star, but may lack additional information needed to confirm it as an exoplanet. Confirmed and statistically validated planets are also in the class PC.  The class of AFP includes eclipsing binaries, background eclipsing binaries and unresolved contamination from other astrophysical sources.  We drop undispositioned \Kepler objects of interest from the training set. Eight of the PCs in the Q1-Q12 public Kepler objects of interest catalog maintained by the NASA Exoplanet Archive have been corrected to AFPs with information that was available at a later date.

After ephemeris matching our training set has 2,879 PCs and 393 AFPs. This likely underrepresents the true population of AFPs since many AFPs were excluded from our pipeline run (see section \ref{sec:tceCatalog}). 

\subsection{Priors}
The random forest algorithm is not a Bayesian technique.  Its knowledge of $P(y_i)$ is inferred directly from the frequencies of the classes present in the training set not through assumption (or knowledge) of some continuous probability function.

$P(AFP)$ when computed this way is problematic in this dataset since this class is underrepresented in the training data (see section \ref{sec:tceCatalog}).  So the estimate of $P(AFP)$ is likely to underestimate the true $P(AFP)$.   

$P(NTP)$ is highly dependent on the algorithm used to detect TCEs; the periodic, instrumental noise present; and the population of target stars. The most comprehensive alternative attempt at discovering planets in the \Kepler data is detailed in \citet{petigura2013prevalence}. In this alternative pipeline 16,227 TCEs were detected.  So it would seem transit detection algorithms lead to an abundance of false detections.

$P(PC)$ conflates the detectability of planets with the actual occurrence rates of planets within our galaxy.  Detectability is most influenced by the the signal-to-noise ratio which is proportional to the square root of the number of transits.  Section \ref{sec:longer} partially addresses the signal-to-noise issues.  We discuss how the random forest fares when the training data is restricted to shorter period TCEs, but is asked to predict against longer period TCEs.  The \Kepler team is still in the process of assessing the detection rate of the pipeline.

Our approach to dealing with priors is to take the frequency of different classes as is and then to reweigh the votes of the random forest to account for the effects of an underrepresented frequency of astrophysical false positives.  Sections \ref{sec:later}  and \ref{sec:betterPriors} discuss how the classifier fares with these corrected priors on an expanded data set with more known classifications.

\subsection{Matching Ephemerides}
\label{sec:matchingEphemerides}
We match the ephemeris contained in the \Kepler object of interest catalog to the TCE catalog in order to identify labels for training data.  Additionally, this algorithm is also used to compute correlations with other TCEs.  This turns out to be an important attribute for class AFP and NTP (section \ref{sec:attributes}).

For each \Kepler object of interest $ K_k$, we identify all the TCEs $ J_j $ where $C_k = C_j $ (the set of TCEs and Kepler objects of interest that are on the same target star).  We identify the best match to the \Kepler object of interest in the following way.  We compute the Pearson's correlation coefficient between the \Kepler object of interest transits and the TCE transits\footnote{This is literally a sequence of binary zeros and ones.  Zero for out of transit and one for in transit.} . We construct a discrete transit indicator vector $z_k$ and $z_j $ over the observing window of each \Kepler object of interest.  The transit indicator vector takes a value of one during a transit and zero outside of a transit. To compute it, we sample the observing window every 1 minute, which gives a large number of samples during a transit. The sampling time can be made smaller at the expense of computation time.

Each transit indicator vector is normalized by the square root of its sum, so that its dot product with itself is unity.  The correlation for the \Kepler object of interest and the best-matched TCE 
\begin{equation}\label{eqn:matchCorrelation}
    \rho_{match}  = \frac{z_k}{\sqrt{\sum z_k}} \cdot \frac{z_j}{\sqrt{\sum z_j}},
\end{equation}
is the dot product of their normalized transit indicator vectors.

If a transit occurs during a data gap, such as a spacecraft safe mode, it will be missed in the raw data. For the purpose of ephemeris matching, we want to represent all the transits that could have occurred during the observing window, so we adjust the relative epoch to correspond to the first transit that could have occurred after the start of the observing window, in case the first transit was missed due to a data gap. For each \Kepler object of interest $ K_k \in \mathbf{K} $, and each TCE $ J_j \in \mathbf{J}$, we replace the epoch with the relative epoch, which is the time relative to the start of the observing window.

\begin{figure}
\epsfig{file=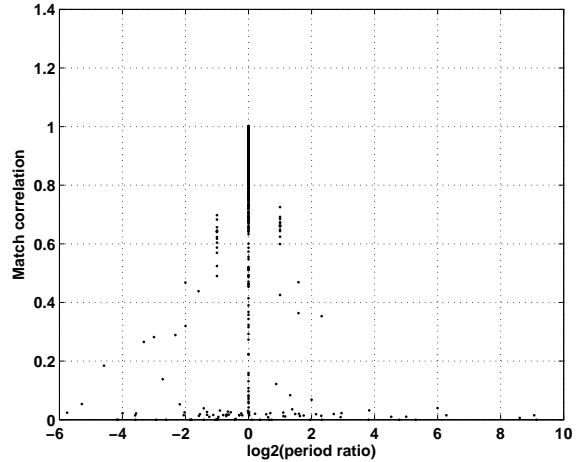, width=3.4in}
\caption{\Kepler object of interest to TCE correlation function vs. \Kepler object of interest to TCE period ratio. The abscissa is $log_2$ of the ratio of the period of the \Kepler object of interest to the period of the best-matching TCE.  The ordinate is the correlation.}
 \label{fig:correlationVsPeriodRatioScatter}
\end{figure}

\begin{figure}
\epsfig{file=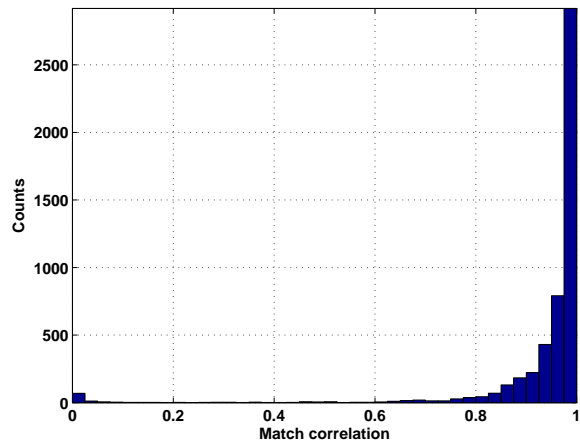, width=3.4in}
\caption{A histogram of the match correlation, showing that the vast majority of \Kepler object of interest matches have correlation near unity.} 
\label{fig:koiTceCorrHist}
\end{figure}
 Evidently from figure \ref{fig:correlationVsPeriodRatioScatter}, a minimum correlation of 0.75 insures that the matched period will never be off by a factor of two.  Accordingly, if the $\rho_{match}$ of the best-matched TCE is greater than 0.75, we accept it as a true match.  If the best match to a \Kepler object of interest has correlation below 0.75, the TCE is labeled as unknown and not included in the training set.

\section{Attributes}
\label{sec:attributes}

\subsection{Missing Attribute Values}
In some cases attribute values are not available.  This can occur because information is missing in stellar catalogs, the data validation fit fails to converge, or a processing timeout is reached.  Several methods of dealing with missing data were investigated: using sentinel values, imputing missing values and training different classifiers for each subset of missing data.   Imputing missing values involves substituting reasonable values for the missing data.  During training, class-conditional values of mean, median and most common are used.  During prediction the non-conditional version of those functions are used to fill values for the unclassified TCEs.  Training multiple classifiers involves identifying those subsets of attributes that are commonly defined.  A sentinel value is a special value that is not commonly seen in the distribution of the attribute; we use -1.  Using a sentinel value turns out to be the simplest method with the lowest error rate.  Internally, when stellar parameters are not available, data validation uses the sun's parameters.  For example, when stellar radius is not available data validation assumes the unknown radius is 1 $R_{\astrosun}$.

\subsection{Attribute Pruning}

\subsubsection{Introduction}
The initial attribute set contains 237 attributes that are based on the wavelet matched filter used by TPS, transit model fitting, difference image centroids and some additional tests.  

The principle of parsimony says we should prefer a classifier that uses fewer attributes to one that uses more attributes.  Also, should many attributes have poor strength, that is poor prediction value, then their presence in the training set will cause degradation of classification performance.  Finally, attributes that are correlated with each other can cause the random forest's error bound to be weaker than it might otherwise be (equation \ref{eqn:bounds}).  For these reasons we remove attributes from our initial training set.

\subsubsection{Importance}
\label{sec:importance}

An importance function returns an estimate the importance of an attribute relative to other attributes, $ Q : \mathbf{x}_i \to \mathbb{R} $.  Attributes can be ranked by their importance; a measure of how much the attribute influences the error rate of the random forest.  This is computed by using the out-of-bag data, permuting the values of the attribute among the out-of-bag TCEs and recomputing the error rate for the tree.  The importance is then the mean increase in error over all the trees in the random forest when the value of the attribute has been permuted.  Random permutations are used so that values come from the same distribution, but are no longer correlated with any classes.

Sometimes we want to refer to the class-conditional importance $Q(\mathbf{x}_i | \mathbf{y}_j) \to \mathbb{R}$; that is we want to measure how important a variable is to predicting a particular class rather then the importance over all classes. When comparing the importance of an attribute with other attributes we use the maximum of all the class conditional importances for each attribute.  This is represented as $max_c(Q(\mathbf{x}_i| \mathbf{y}_c))$.  In this way an attribute that is important for discriminating only one class will not be discarded in section (\ref{sec:removingWeakAttributes}).

\subsubsection{Removing Weak Attributes}
\label{sec:removingWeakAttributes}
In order to identify weak attributes we must first know what the importance of a weak attribute looks like.  To do this we add an attribute that is generated with a random number generator to the training set and then measure the importance of this random attribute. One hundred and fifty one random forests were trained, each with a different randomly generated attribute. This gives us an estimate of the importance of a random attribute, $Q(\mathbf{x}_{random})$.  We choose a cutoff of $ 6 \sigma $ of $Q(\mathbf{x}_{random})$ above zero importance; this turns out to be 5E-5.  Attributes that have a maximum, class-conditional importance less than this threshold are dropped.  This removes 35 attributes.

\subsubsection{Removing Correlated Attributes}
\label{sec:removeCorrelation}
Equation \ref{eqn:bounds} implies that removing correlated attributes can decrease the error rate of the random forest as it will decorrelate errors.  To remove correlated attributes, we compute the matrix of absolute values of Pearson's correlation coefficients, $ \rho_{i,j} $,  from our matrix of attributes. $\theta_{\rho}$ is a correlation threshold, attributes $ (\mathbf{x}_i, \mathbf{x}_j) $ whose $\rho_{j,j}$ is greater than this threshold are considered significantly correlated.  

For each pair of attributes $ (\mathbf{x}_i, \mathbf{x}_j)$, if $ \rho_{i,j} >= \theta_{\rho}$ then we drop one or the other attribute.  The attribute that has the lower of the two class-conditional importances $max_c(Q(\mathbf{x}_i|\mathbf{y}_c))$ is dropped. 

To estimate $ \corrthreshold$ we trained a random forest by imposing various values of $\corrthreshold \in [0.15,1.0] $ in $0.05$ increments.  We verified that the out-of-bag error (section \ref{sec:randomForestDescription}) differs by only 5E-4 in the interval $ \corrthreshold [0.35,1.0]$.  Removing one of a pair of attributes with lower values of $\theta_{\rho}$ increase the out-of-bag error.  Therefore reducing the attribute correlation threshold does not produce any significant decrease in misclassification error.  However, this still allows us to reduce the number of attributes used by the random forest to produce a more parsimonious training set.

At the minimum out-of-bag error, $ \corrthreshold = 0.55$ and there are 77 attributes that meet our thresholds. The ten most important of these attributes are described in detail in section \ref{sec:mostImportantAttributes}.

\subsection{Most Important Attributes}

\label{sec:mostImportantAttributes}

We discuss the top 10 most important attributes in this section.  This cutoff is arbitrary for the sake of brevity.  The range of attribute importance of the top 10 attributes is from 7.5\% increase in out-of-bag error down to a 5.3\% increase in out-of-bag error.  For the purpose of ranking attributes by importance we use the maximum of the class-conditional importances.

\begin{figure*} 
\epsfig{file=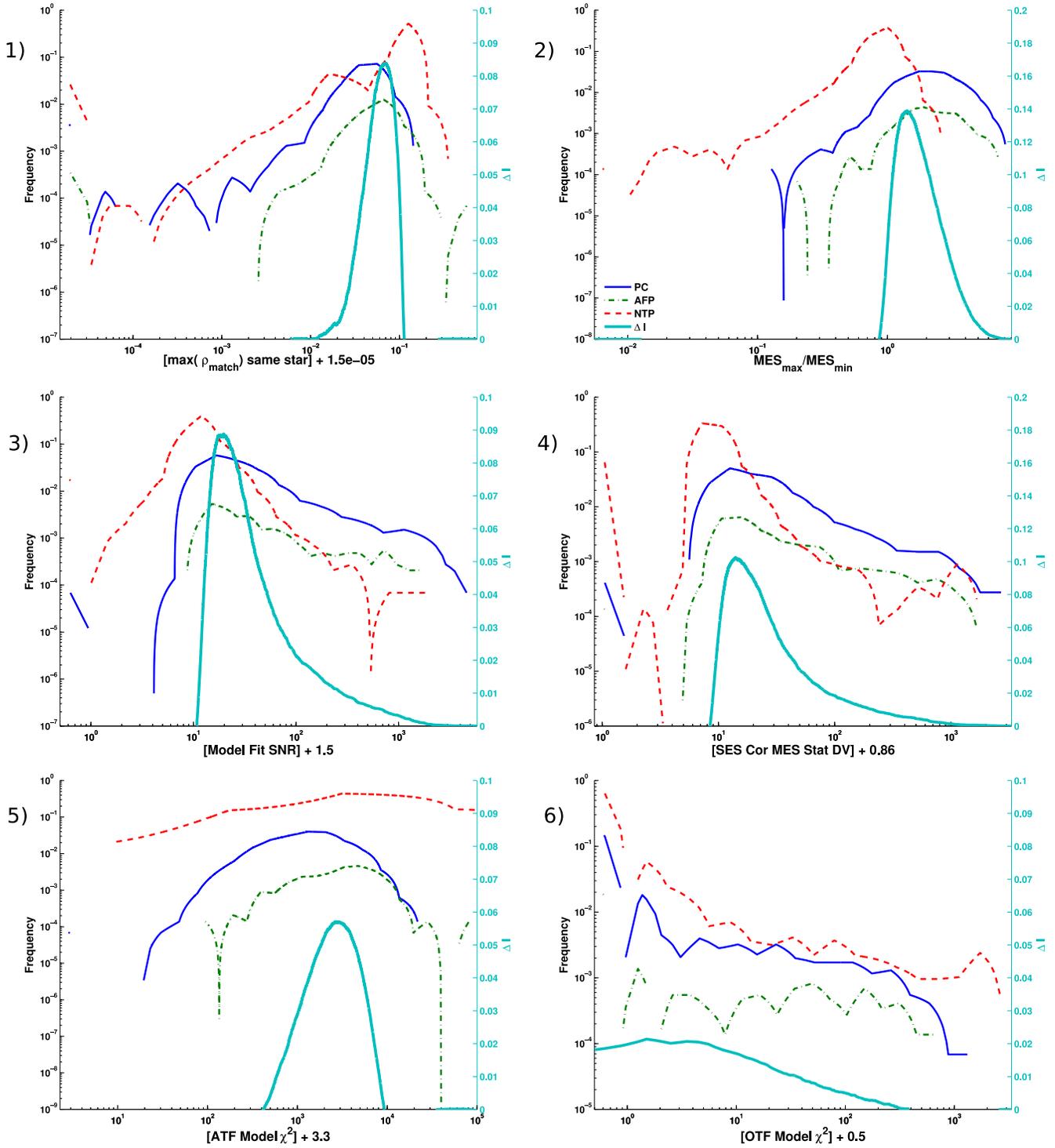,width=7.0in}
\caption{Class-conditional distributions of attributes 1 through 6.  If an attribute's domain extends to zero or negative numbers a small constant has been added so that it can be log scaled.  $\Delta \impurity$ (equation  \ref{eqn:impurityDelta}) is plotted to show where a split might be made if it was the first split in a decision tree.}
 \label{fig:distributions1of2}
\end{figure*}

\begin{figure*} 
\epsfig{file=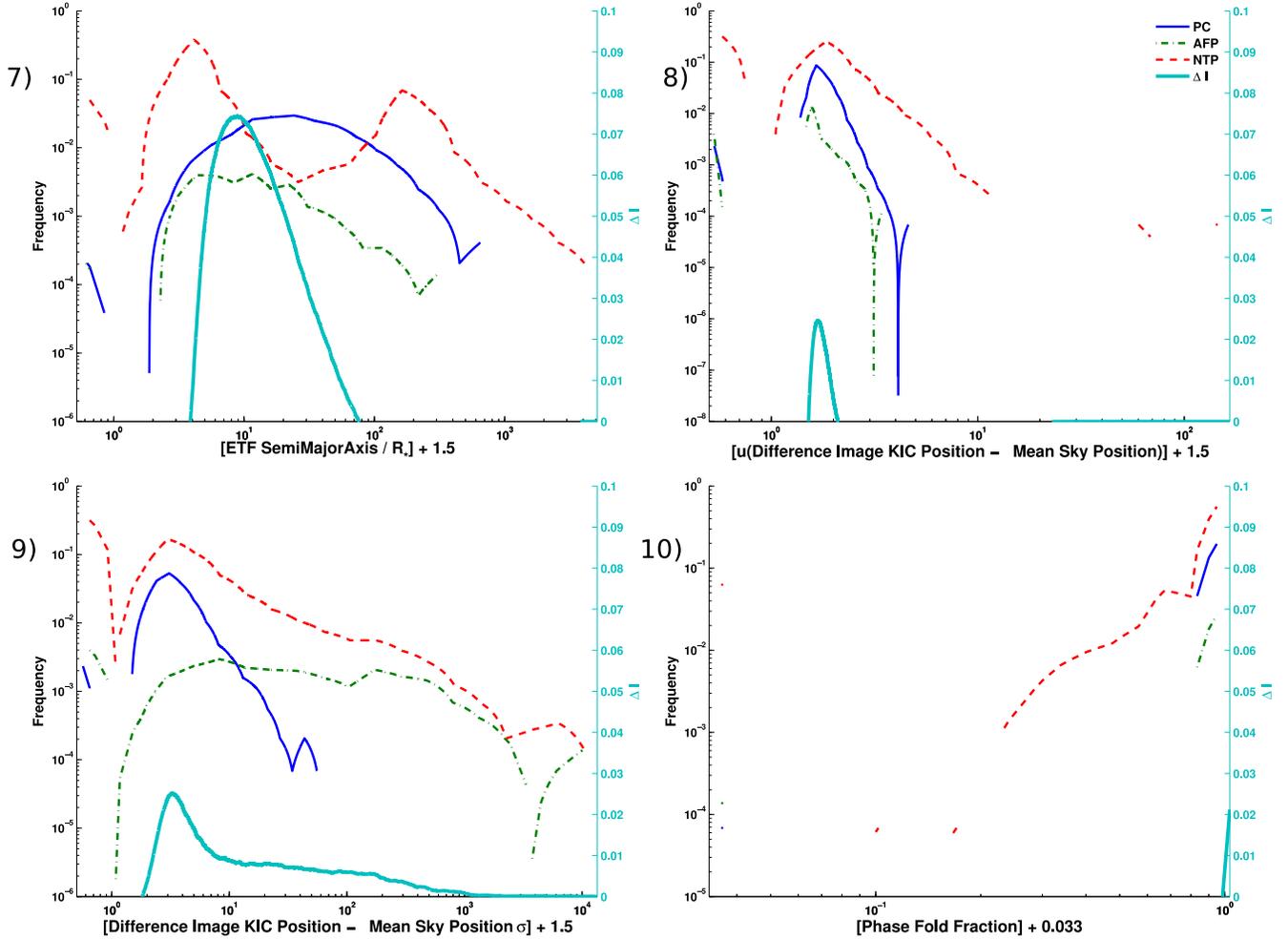, width=7in}
\caption{Class-conditional distributions of attributes 7 through 10.  If an attribute's domain extends to zero or negative numbers a small constant has been added so that it can be log scaled.   $\Delta \impurity$ (equation  \ref{eqn:impurityDelta}) is plotted to show where a split might be made if it was the first split in a decision tree.}
 \label{fig:distributions2of2}
\end{figure*}

Figure \ref{fig:distributions1of2} and figure \ref{fig:distributions2of2} have the class-conditional, attribute distributions for these attributes.

\begin{enumerate}
\item \label{attr:maxEphemerisCorrelationSameStar} The maximum ephemeris correlation, $\rho_{match}$, (equation \ref{eqn:matchCorrelation}), is computed for a TCE against all the other TCEs on the same star.  If there is only one TCE on the same star then this value is zero.  When the ephemerides of TCEs on the same star are highly correlated it is an indication that a secondary eclipse may be present and has been detected as a TCE.  For NTPs this may indicate that a similar source of instrumental noise has been detected with a slightly different phase.

\item \label{attr:allPulseMaxMesMinMesRatio} The absolute value of the ratio of the maximum MES to the minimum MES (see section \ref{sec:TPS}) found over all search periods.  When this value is near 1 it indicates that the periodic events lower than the median flux are of similar statistical significance to similar events greater than the median flux.  This attribute attempts to eliminate TCEs that are problematic with respect to stellar variability and instrumental noise.

\item \label{attr:ATFModelFitSnr} The SNR for the all-transit model fit.  This is the transit depth normalized by its uncertainty.

\item \label{attr:sesCorMesStatDV} MES scaled by SES auto-correlation statistic; $MES (1 - R)$.  The expected SES time series should have negligible auto-correlation strength, $R$, beyond the transit duration time scale.  This expectation is not always met in practice for \Kepler flux time series having residual astrophysics and systematics remaining after the detrending and whitening filter is applied.

To identify targets with enhanced auto-correlation strength beyond the expectation, we begin with the SES time series assuming a 10.5 hour transit duration and remove cadences with bad data.  The remaining good cadences are separated into blocks of 4,250 cadences (similar to a single \Kepler observing quarter).  For each block of 4,250 cadences the auto-correlation is calculated out to lag 1050, and we calculate the fraction of the lags where the auto-correlation value is above or below the 95\% or 5\% expectation, respectively.  We simulate the expected auto-correlation value through 150k Monte Carlo trials.  A Monte Carlo trial starts with a random realization of a zero-mean, unit-variance Gaussian vector of length 4,250 with a moving average applied with a 10.5 hour window size.  The Monte Carlo trial roughly corresponds to the expectation for the SES time series in the null case.  We measure the auto-correlation on the Monte Carlo trials to empirically estimate the 5\% lower and 95\% upper bound expectation at each lag.

The expectation for a null result (well behaved) SES time series is an auto-correlation strength of 0.05; meaning 5\% of the lags for a typical SES time series are above the expectation set by the Monte Carlo simulation.  As values approach 1.0 for the auto-correlation strength, the SES is demonstrating very large amounts of auto-correlation and can violate an underlying assumption of the  detection statistics.  We use the median of the auto-correlation strength across all the blocks to scale the detection statistic, $R$.

\item \label{attr:ATFModelChiSquare} The $\chi^{2}$ for the all-transit model fit.  This is the raw $ \chi^{2}$ value; that is, not scaled by the degrees of freedom.

\item \label{attr:OTFModelChiSquare} Same as attribute \ref{attr:ATFModelChiSquare}, but this only considers the odd numbered transits.  Comments about odd and even transits made in attribute \ref{attr:ETFRatioSemiMajorAxisToStarRadiusV} apply equally here as well.   The Kolmogorov-Smirnov test shows the importance of this attribute is significantly greater than the all-transit-fit version of this attribute (attribute \ref{attr:ATFModelChiSquare}) for class NTP with $ p \ll 0.001$.    

\item \label{attr:ETFRatioSemiMajorAxisToStarRadiusV}The ratio of the fitted planet's semi-major axis to the star radius for only the even numbered transits.

Sampling only the even transits (or conversely the odd transits) may have the effect of skipping or including quarters of data that are more problematic.  During the second quarter of operation \Kepler experienced a number of safe modes causing thermal transients.  During the fourth quarter one of the CCDs modules failed.  During the eighth quarter a number of safe modes occurred.  \Kepler experienced a coronal mass ejection event during the twelfth quarter.

In order to see if the importance of this attribute was due to the randomness of the random forest we trained 40 random forests and then tested the distributions of the importances for this attribute vs the distribution of the importances of the all-transit-fit version of this attribute.  The Kolmogorov-Smirnov test shows the importance of this attribute is significantly greater than the all-transit-fit version of this attribute for class NTP with $ p \ll 0.001$.  

Additionally we computed the importance of this attribute for every TCE.  The TCEs that show this attribute as having higher importance than all-transit-fit version of this attribute tend to have longer periods; greater than 4 days.  This also corresponds with the somewhat bimodal distribution of NTPs which tend to be more prevalent at relatively shorter and longer periods as shown in plot of attribute \ref{attr:ETFRatioSemiMajorAxisToStarRadiusV} in \ref{fig:distributions2of2}.  There is a weak negative correlation of -0.22 (Pearson's correlation coefficient) with the effective temperature of the target star.

\item \label{attr:DIMRMqKicCentroidOffsetsMeanSkyOffsetU} The uncertainty of the angular offset on the plane of the sky between the best-fit centroids from the difference image and the \Kepler input catalog (KIC) position by averaging over all quarters.  The KIC position is subtracted from the difference image centroid.  TCEs with large offsets from the KIC position are believed to be caused by background eclipsing binaries or some other events that are not on the target star. 

\item \label{attr:DIMRMqKicCentroidOffsetsMeanSkyOffsetS} 
The significance of attribute \ref{attr:DIMRMqKicCentroidOffsetsMeanSkyOffsetU} rather than the uncertainty itself.

\item \label{attr:phaseFoldLCFraction} The proportion of the light curve that was missing during this TCEs transit.  We use an already detrended light curve that is median detrended with a window 3 times the transit duration.  This is folded on the TCE ephemeris. The detrended light curve is binned with a bin width of half a transit duration for bins within 4 transit durations of the transit center. This attribute is the fraction of the bins that have any data contributing to them. A value of less than one indicates the TCE is overlapping with a region of data that are missing when phase folded on the ephemeris.

\end{enumerate}

\section{Classification Results}
\label{sec:results}
\subsection{Assessment}
When assessing classification performance we use the out-of-bag error (section \ref{sec:randomForestDescription}) rather than using cross validation (the common approach for estimating the misclassification rate).  We tuned the number of trees in the random forest so that the out-of-bag errors are close to the leave-one-out-cross-validation votes.  When $ N_{trees} = 1,000$ out-of-bag votes have a correlation with the leave-one-out-cross-validated votes of 0.90.  When $ N_{trees} = 3,000$ this correlation becomes 0.998.  We use $ N_{trees} = 15,000$, since this allows us some margin of error and is still computationally tractable.  Since each TCE used for training is out of bag a random number of times the votes are normalized by the number of trees for which the training TCE was out-of-bag so that $\sum\limits_{i=1}^{C} v_i = 1 $.   

The number of attributes considered for any split of the training set is known as $m_{try}$ (section \ref{sec:randomForestDescription}).  The value of $m_{try}$ suggested by \citet{liaw2002Random}  is $\lceil \sqrt{A}\rceil$, where $A$ is the number of attributes.  In practice the random forest is not very sensitive to changes in $m_{try}$.  For our final 77 attributes $m_{try} = 9$.  

We are able to achieve out-of-bag error rate of  2.19\%.

\subsection{Optimal Loss Function}
 It's possible to stop here and use this vote tuple to assess the performance, but we also weigh the votes so they minimize the out-of-bag error.  We do this in a manner that allows us to change the our assumptions about the priors.
 
We compute the confusion rate matrix.  The columns of a confusion matrix indicate the predicted classes, while the rows indicate the acutal class. An entry in the diagonal indicates the count correct classifications, off-diagonal entries indicate misclassifications of various kinds.  For the ideal confusion matrix the sum of the diagonal elements is equal to the number of items in the training set.  A confusion rate matrix is a rate rather than a count of classifications.  For the ideal confusion rate matrix the diagonal would be the frequency of the classes present in the training set.
 
The out-of-bag votes are  $v_i(l)$ for class $y_i$ for TCE $l \in \mathcal{L}$.  The votes are weighted by $\Omega = { \omega_1, \omega_2, \omega_3} $.  The function $I: y \to {0,1}$, is an indicator function that returns 1 for correct classifications else it returns 0.  Parameter $ \omega_1 $ is not a free parameter and is set to 1.  The confusion rate matrix is defined as 
\begin{equation} \label{eqn:confusionRateMatrix}
\zeta_{i,j}(\Omega) = \frac{1}{|y_i|} \sum\limits_{l \in \mathcal{L}} I\left(y_j = \argmax_{i=1}^{C} \{\omega_i v_i(l)\}\right).
\end{equation}

We followed the methodology of \citet{landgrebe2007approximating} and chose an $\Omega$ the minimizes following the loss function representing the Bayes costs for the given set of weights:
\begin{equation} \label{eqn:bayesLoss}
\begin{split}
L(\Omega)  = &\sum\limits_{i=1}^{C} P(y_i)\bigg(\sum\limits_{j=1,j \neq i}^{C}\zeta_{i,j}(\Omega) S_{i,j}\bigg) - \\
    & \sum\limits_{i=1}^{C} P(y_i) \zeta_{i,j}(\Omega) S_{i,j},
\end{split}
\end{equation}
where $S_{i,j}$ is the cost matrix, and $\zeta_{i,j}$ is the confusion rate matrix specifying the fraction of objects with label $i$ classified as class $j$ by the classifier with class weights $\Omega$.  In this paper we set the cost matrix to have zeros along its diagonal and ones elsewhere.  This corresponds to minimizing the total number of misclassifications.  A logarithmically spaced grid is used to search for values of $\Omega$ that minimize equation \ref{eqn:bayesLoss}.   Prior probabilities, $P(\mathbf{y}_i)$, of classes are estimated from the frequency of the classes in the training set.
  
At the optimal values of $\Omega$ the random forest is able to achieve out-of-bag error of 1.34\%, an improvement of 0.85\% over the unweighted version. The resulting confusion matrix is presented in table \ref{table:rfConfusionMatrix}.  The diagonal elements of a confusion matrix represent correct classifications.  Off-diagonal elements are errors.  The error rate can be computed from the confusion matrix by summing all the off-diagonal elements and then dividing by the sum of all matrix elements. s
\begin{table} 
\centering
\caption{Random forest confusion matrix}
\label{table:rfConfusionMatrix}
\begin{tabular}{c c c c}
Training Set Class &  \multicolumn{3}{c}{Predicted Class} \\
                   & PC    & AFP    & NTP    \\
\hline 
        PC         & 2,843  & 8      & 28     \\
        AFP        & 98    & 271    & 24     \\
        NTP        & 25    & 12     & 11,267  \\
\end{tabular}
\end{table}

\begin{figure} 
\epsfig{file=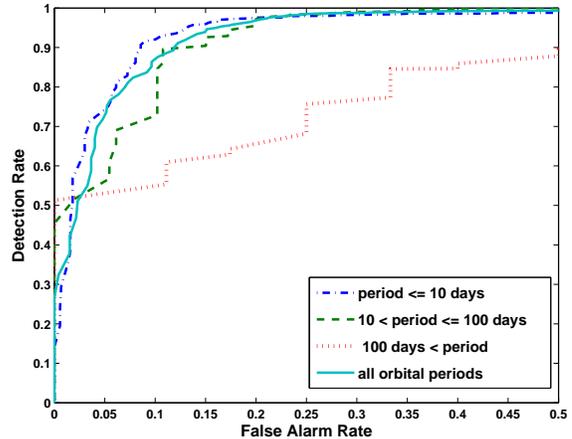,width=3.4in}
\caption{Period dependent ROC curve for PC vs AFP. For all orbital periods the AUC is 0.9522 $\pm$  0.0080 .  For the orbital period bins (p $\le$ 10, 10 $<$ p $\le$ 100, and p $>$ 100) the AUC are 0.9568 $ \pm $ 0.0094, 0.9484 $\pm$ 0.0193, and  0.8229 $\pm$ 0.0730 respectively.}
\label{fig:periodDependentPCvsAFP}
\end{figure}

From the weighted out-of-bag votes we produce a Receiver Operating Characteristic (ROC) curve that shows detection rate as a function of false positive rate.  To generate the curve we calculate the false positive rate at vote thresholds in .01 increments in the interval [0,1].  A vote threshold of one is accepting a classification only when all trees are unanimous and a zero vote threshold is accepting every classification for the class under examination.  A vote threshold of zero indicates that an instance of the class will never be misclassified, but guarantees a high false positive rate.  The detection rate (DR) is the number of instances of the class correctly classified divided by the number present in the training data.  The false alarm rate (FAR) is the number of misclassifications divided by the number of instances of the undesirable classes detected in the training set.  The Area Under the Curve (AUC) is the integrated area under the ROC curve.  This is a useful metric when comparing classifiers.  A classifier with a superior AUC will have some trade off between DR and FAR that results in higher DR/FAR ratio than one with a lower AUC.  For more information on the receiver operating characteristic see \citet{fawcett2004roc}.
Planet candidates vs astrophysical false positives show an orbital period dependent ROC curve, see figure \ref{fig:periodDependentPCvsAFP}.  This is likely caused by a lower detection rate for \Kepler at longer periods and therefore less training data at longer periods for all classes.

\subsection{Comparison With Other Classification Algorithms}
In addition to the random forest we also tried other well-known classification algorithms that are often used as benchmarks for other classification problem domains: k-nearest neighbors (k-NN) and naive Bayes.

\begin{figure}
\epsfig{file=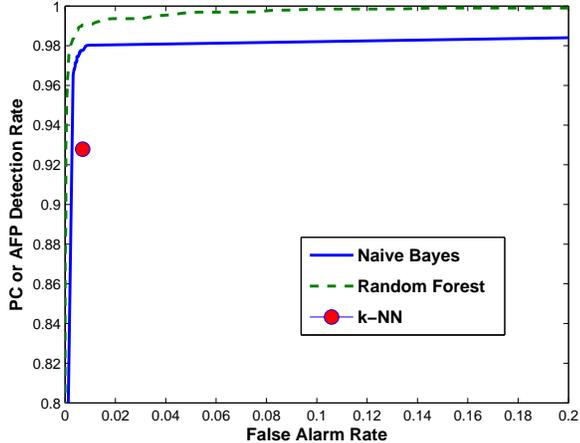,width=3.4in}
\caption{Comparison of random forest, naive Bayes and k-NN. AUC=0.9991 for random forest and  0.9894 for naive Bayes. k-NN (when $k=1$) does not produce a ranking of predictions and so is represented as a single point in the trade space between detection rate and false alarm rate.}
 \label{fig:classifiersCompared}
\end{figure}

The k-NN algorithm \citep{duda2001Pattern} has the advantage that it is simple, explainable and provides bounded error, but in a different manner to the random forest.  With training set of infinite size k-NN would converge to no more than twice the error bounds of a Bayesian classifier with perfect knowledge of the joint probability distribution function of all the attributes. k-NN classifies an unknown using the minimum distance to some example in the training set.  The Euclidean distance function between points in the attribute space is used. The notion of distance in can be broadened to include the class of the $k$ nearest neighbors instead of just a single nearest neighbor.  For our problem this turns out not to be useful; optimal $k$ turns out to be $k = 1$.  Other values of $k$ turn out to have classification performance very close to random.  An identical training set is used for RF and k-NN except that the attributes are log scaled. To assess k-NN performance we use leave-one-out cross validation.  Unfortunately, since $k = 1$, it's not possible to plot an ROC curve for k-NN since its classification space is just a single point (figure \ref{fig:classifiersCompared}).

A naive Bayes classifier assigns a classification by choosing the class with the highest probability given the evidence.  This algorithm has the advantage that it is simple, explainable and provides probabilities rather than votes.
\begin{equation} \label{eqn:naiveBayes}
\begin{split}
 i^* = & \argmax_{i} P(\mathbf{y}_i| x_u) \\
P(\mathbf{y}_i|x) = & P(\mathbf{y}_i) \prod\limits_{j=1}^{A} \frac{P(x_j|\mathbf{y}_i)} {P(x_j)}
\end{split}
\end{equation}
Where $x$ refers to the attributes of the unknown TCE and $A$ is the number of attributes.  Prior probabilities, $P(\mathbf{y}_i)$, of classes are estimated from the frequency of the classes in the training set.  Probability distributions, $P(x_j|\mathbf{y}_i)$, are estimated using a kernel density estimator.  We use a Gaussian kernel with Silverman's rule-of-thumb bandwidth described in \citet{feigelson2012Modern}.  The implication of equation \ref{eqn:naiveBayes} is that our attributes are independent from one another;  it is naive after all.

We use leave-one-out cross validation to estimate the classification performance.  The resulting error rates for k-NN and naive Bayes are 3.15\% and 2.73\% respectively.  Performance of all classification algorithms described in this paper are presented for the purpose of discriminating between  $ PC \cup AFP$  vs  $ NTP $ in figure \ref{fig:classifiersCompared}.

These results show that while other classification algorithms have better than random performance they do not perform as well as the random forest. 
 
\FloatBarrier

\section{Application to Transit Surveys}
\label{sec:application}
\subsection{Classifying Against Longer Orbital Periods Than Training Data}
\label{sec:longer}
A transit survey such as \Kepler will initially discover larger, shorter period planets.  As the survey continues it will find smaller, longer period planets.  This poses a potential issue to supervised machine learning algorithm as a portion of parameter space it is trained on may no longer be representative of the population as a whole.  We look at what happens to classification errors when we only have successively smaller amounts of training data and for shorter orbital periods.  

\begin{figure}
\epsfig{file=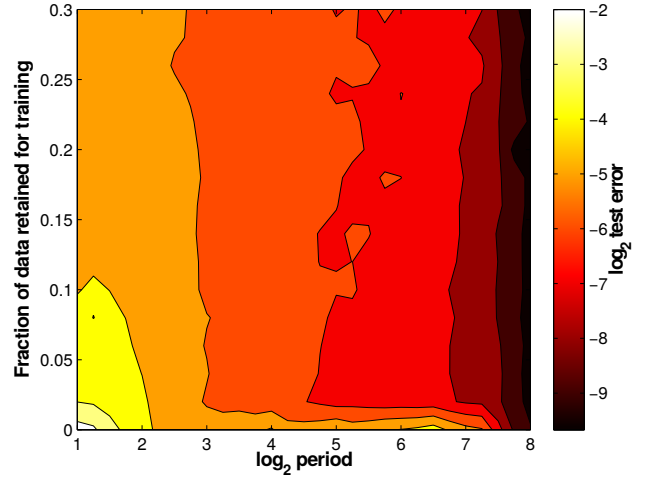,width=3.6in}
\caption{The effects of limiting the amount of training data used to less than some orbital period.  The abscissa is the $log_2$ orbital period.  The ordinate is the portion of the TCEs retained for training that have orbital periods less than the abscissa.  The value plotted is $log_2$ of the misclassification error of all the data that was not used for training.}
\label{fig:observingTimeDependent}
\end{figure}

Figure \ref{fig:observingTimeDependent} shows the error rates achieved when limiting the training data used at various orbital period thresholds. The error is computed by training three random forests for each combination of period and threshold evaluated.  The error for an individual random forest is then computed on the test set (the portion of the training data that was not used for training) and the out-of-bag error.  The mean out-of-bag error of the three random forests is used to estimate the error surface in figure \ref{fig:observingTimeDependent}.

At a period of 16 days and training data fraction of zero no training data longer than 16 days is used to train the random forests.  The error rate for this orbital period is ~ 3.13\%.  If we were then expanding our search to longer periods then  we would only need to add  5\% of the TCEs with periods longer than 16 days to have the error rates fall to ~ 1.56\%.   Therefore adding a small, but uniformly sampled (with respect to orbital period) portion of the unknown TCEs to the training set can yield a large reduction in misclassification error.

\subsection{Predictions on Later TCERT Dispositions}
\label{sec:later}
The TCE catalog contains 3,831 TCEs that are not labeled with our ephemeris matching algorithm.  Of these, 1,487 have since been given dispositions by the TCERT using somewhat different techniques from the original training set using methods described in \citet{coughlin2014Contamination}.  We look at the classification predictions on this later set of TCEs that were not in the training set.  Table \ref{table:predictionConfusionMatrix} contains the confusion matrix of the random forest predictions.

\begin{table} 
\centering
\caption{Random Forest Predictions vs Later TCERT}
\label{table:predictionConfusionMatrix}
\begin{tabular}{c c c c}
TCERT Class & \multicolumn{3}{c}{Predicted Class} \\
                     &  PC     & AFP     & NTP \\
\hline                                        
         PC          &  336    & 10      & 43 \\
         AFP         &  204    & 388     & 506 \\
\end{tabular}
\end{table}

We looked at a random sample of a small number of AFPs that were misclassified as PCs.  The vast majority of these were determined by the TCERT to be AFPs based on information not available in the training set of attributes.  These classifications are based on the presence of significant secondary eclipses, indicating an eclipsing binary.  Another source of false positives is direct PRF contamination.  In these cases the light from a nearby, but out-of-aperture star, is inducing a transit signal on the target star and our existing attributes are not able to detect the off aperture false positive. 

AFPs that are misclassified as NTP are partially due to a labeling problem that the TCERT will remedy in the future: the \Kepler object of interest table generated by TCERT homogenizes all kinds of non-planetary TCEs into one category: false positive.  Non-planet TCEs that are astrophysical in nature and those that are instrumental both receive the same label, false positive.  This is why there is not a NTP row in table \ref{table:predictionConfusionMatrix}.  

\subsection{Using Corrected Priors}
\label{sec:betterPriors}
The priors used in our training set are much different compared with the test set used in section \ref{sec:later}.   Using equation \ref{eqn:bayesLoss} we can reweigh the votes of the trained classifier to correct for the eclipsing binaries omitted from the run of the \Kepler pipeline and so do not appear (see section \ref{sec:tceCatalog}).  Table \ref{table:predictionCFWeighted} shows improvement when more realistic priors are used.  Table \ref{table:totalCF} is the combined confusion matrix representing the TCEs in table \ref{table:predictionCFWeighted} and table \ref{table:rfConfusionMatrix} (with different priors).

With priors reflecting more eclipsing binaries the total error from table \ref{table:totalCF} is 5.85\%.  When the errors between AFPs and NTPs are ignored we have an error rate of 2.81\% which is more relevant for planet detection.

\begin{table} 
\centering
\caption{Random Forest Predictions vs Later TCERT, Corrected Priors}
\label{table:predictionCFWeighted}
\begin{tabular}{c c c c}
TCERT Class & \multicolumn{3}{c}{Predicted Class} \\
                     &  PC     & AFP     & NTP \\
\hline                                        
         PC          &  314    & 32      & 43 \\
         AFP         &  95    & 553     & 450 \\
\end{tabular}
\end{table}

\begin{table} 
\centering
\caption{Total, Prior-Corrected}
\label{table:totalCF}
\begin{tabular}{c c c c}
Training Set Class &  \multicolumn{3}{c}{Predicted Class} \\
                   & PC    & AFP    & NTP    \\
\hline 
        PC         & 3,040  & 151    & 77     \\
        AFP        & 147    & 882    & 462     \\
        NTP        & 17     & 35     & 11,252  \\
\end{tabular}
\end{table}

\FloatBarrier

\section{Conclusion}
\label{sec:conclusion}
Machine learning techniques offer a way to automate some stages of exoplanet discovery.  In this paper we have demonstrated that the random forest algorithm is quite good at distinguishing between systematic noise, eclipsing binary  and exoplanet candidate signatures. As seen in table \ref{table:totalCF}, classifications have a low overall error rate (5.85\%), when ignoring distinctions between astrophysical false positives and noise the error rate falls to 2.81\%.

When used to predict on longer orbital periods than the training data the overall error rate increases.  This can be mitigated by sampling a small percentage of the more recent TCE detections, classifying them and adding them to the training set.  There are some populations of astrophysical false positives that have been identified using other methods; with attributes not available in the training data set. The random forest often confuses these astrophysical false positive TCEs with planet candidates and non-transiting phenomena (table \ref{table:predictionConfusionMatrix}).

In the future we intend on adding additional attributes that would help separate out the astrophysical false positives.  Newer versions of the TPS software can detect weak secondary eclipses that would indicate the presence of an eclipsing binary.  Transit signatures induced by off-aperture sources can be identified by performing photometry on the local background pixels of the target stars.  Transits detected in this aperture should then not have similar significance to the transits detected on the target stars.

The \Kepler mission is currently working on estimating the completeness of the pipeline using injected transits \citep{christiansen2013Measuring}.  Using injected transits as training data can will allow us to assess the accuracy of human and machine classification efforts.  We look forward to continued revision and applications of these automated classification methods on subsequent \Kepler pipeline runs, and these methods can be readily applied to large transit surveys in the future such as TESS \citep{ricker2014Tess}.

\pagebreak

\section*{acknowledgments}
\Kepler was competitively selected as NASA's 10th Discovery mission.  We would like to thank Abhishek Jaiantilal for use of the randomforest-matlab code.  This paper would not be possible without the work of the members of the \Kepler TCE Review Team.  Funding for the \Kepler mission is provided by NASA's Space Mission Directorate.  This research has made use of the NASA Exoplanet Archive, which is operated by the California Institute of Technology, under contract with NASA under the Exoplanet Exploration Program.

\bibliography{autovetting}

\end{document}